\documentclass[a4paper]{jpconf}
\usepackage{graphicx}
\begin{document}
\title{JUNO Non-oscillation Physics}

\author{Giulio Settanta, on behalf of the JUNO Collaboration}

\address{Forschungszentrum Jülich GmbH, Nuclear Physics Institute IKP-2. Wilhelm-Johnen-Straße, 52428 Jülich, Germany}

\ead{g.settanta@fz-juelich.de}

\begin{abstract}
The JUNO observatory, a 20\,kt liquid scintillator detector to be completed in 2022 in China, belongs to the next-generation of neutrino detectors, which share the common features of having a multi-ton scale and an energy resolution at unprecedented levels. 
Beside the ambitious goal of neutrino mass ordering determination, the JUNO Collaboration plans also to perform a wide series of other measurements in the neutrino and astroparticle fields, rare processes and searches for new physics. The detector characteristics will allow the detection of neutrinos from many sources, like supernovae, the Sun, atmospheric and geoneutrinos. Other potential studies accessible to JUNO include the search for exotic processes, such as nucleon decays, Dark Matter and magnetic monopoles interactions, light sterile neutrinos production.\\
This work reviews the physics potential of JUNO about non-reactor neutrino sources, highlighting the unique contributions that the experiment will give to the various fields in the forthcoming years.
\end{abstract}

\section{Introduction}
The JUNO observatory, a 20\,kt liquid scintillator (LS) --based experiment currently under construction in China, belongs to the next-generation of neutrino detectors, which share the common features of having a multi-ton scale and an energy resolution at unprecedented levels. Representing the major expression of new technologies in neutrino physics, they are expected to play a primary role in answering the still open issues in the field. JUNO, in particular, has its main goal in the identification of the neutrino Mass Ordering (MO), which has several fundamental implications in Particle Physics and Cosmology. The measurement will rely mainly on the reconstruction of fine structures in nearby reactors antineutrino spectrum. JUNO excellent expected performances come from its design: the detector consists of a layered structure, the innermost part being a 35.4\,m diameter acrylic sphere which contains the LS. The light produced in neutrino interactions with the LS is collected by a double system of photosensors: 17612 20'' PMTs and 25600 3'' PMTs. The central spherical detector is surrounded by a cylinder--shaped active water Cherenkov detector, to reduce the background produced by atmospheric muons. On top of that, a 3--layer tracker, made of plastic scintillator strips, is placed \cite{JUNO_CDR,JUNO_NYB}. The expected energy resolution of the detector is 3\%$\sqrt{E [MeV]}$.

Beside the ambitious goal of neutrino MO identification, the JUNO Collaboration plans also to perform a wide series of other measurements in the neutrino and astroparticle fields. They include neutrinos from supernovae and the Sun, as well as atmoshperic and geo--neutrinos. Signatures of exotic processes, not allowed by the Standard Model of Particle Physics, could be also detected \cite{JUNO_NYB,JUNO_YB}.

\section{Supernova neutrinos}
\begin{figure}[htbp]
\begin{center}
\includegraphics[width=0.55\textwidth]{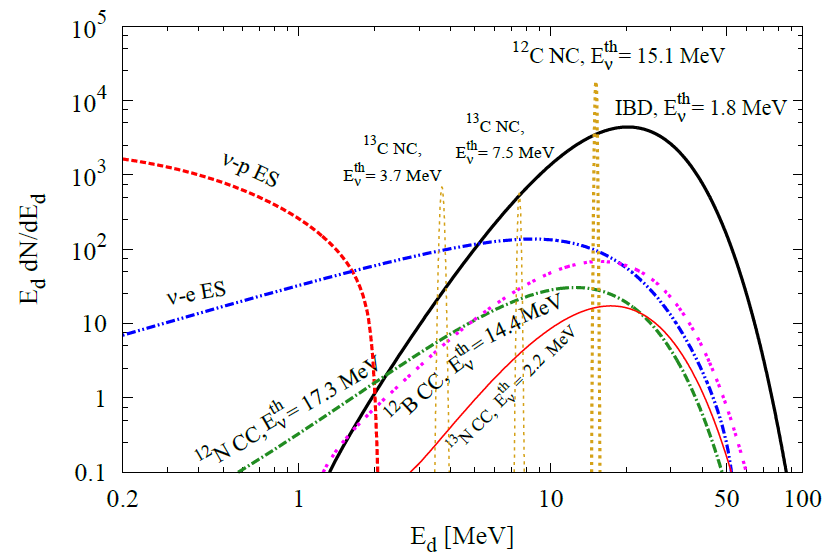}
\caption{Distribution of the visible energy $\mathrm{E_d}$ of supernova neutrinos, according to different interaction channels. Figure from \cite{JUNO_NYB}.}
\label{fig:sn_spectra}
\end{center}
\end{figure}

The large mass of the LS allows the detection of neutrinos of all flavors from a supernova explosion, with high statistics. A JUNO observation of supernovae neutrinos, in the energy range of tenths of MeV, would provide unique information about the explosion mechanism and about neutrino physics itself. Figure \ref{fig:sn_spectra} reports the energy spectra, in terms of the visible energy, of supernova neutrinos in case of a stellar explosion inside our galaxy. Different interaction channels produce different energy distributions, representing a major improvement with respect to the SN1987A observation. Some studies show that JUNO could be sensitive also to pre--supernova neutrinos, representing a powerful tool for an early supernova alert system \cite{Pre_SN}. A crucial role will be played by the JUNO Multi--Messenger trigger, which is under development within the Collaboration.

\begin{figure}[htbp]
\begin{center}
\includegraphics[width=0.78\textwidth]{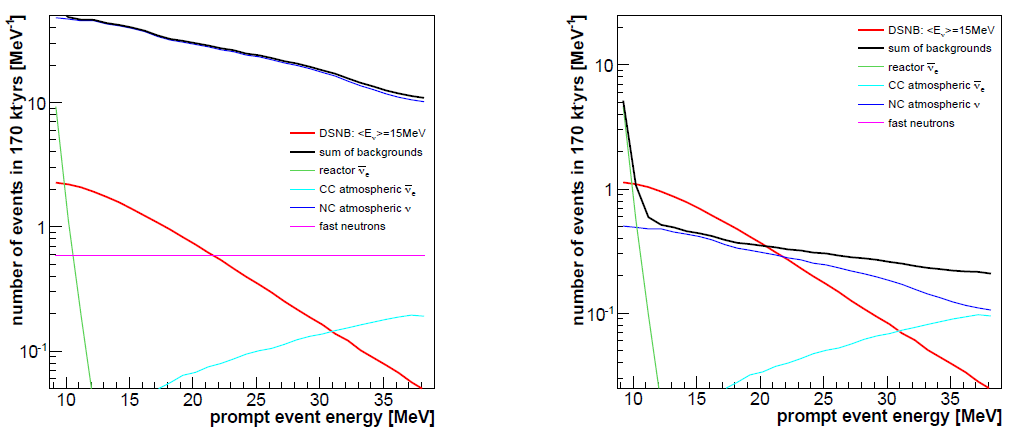}
\caption{Positron energy spectra from DSNB interactions for a 10 years detector exposure, assuming \textless$E_\nu$\textgreater = 15\,MeV (in red), and the expected backgrounds in JUNO. Left: after basic selection cuts. Right: after pulse shape discrimination. Figures from \cite{JUNO_NYB}.}
\label{fig:dsnb_spectra}
\end{center}
\end{figure}

Not only recent supernova explosions could be detected by JUNO, but also the integrated signal from all past ones, namely the Diffuse Supernova Background (DSNB). The large detector mass is again the key point to address this observation, which would return important insights about the cosmological structure. Figure \ref{fig:dsnb_spectra} shows the expected signal from DSNB (after an Inverse Beta Decay interaction) and various backgrounds, assuming a 10 years detector exposure and a mean energy of DSNB neutrinos of 15\,MeV. This last parameter significantly influences the JUNO DSNB sensitivity \cite{JUNO_NYB}.

\section{Solar neutrinos}
\begin{figure}[htbp]
\begin{center}
\includegraphics[width=0.98\textwidth]{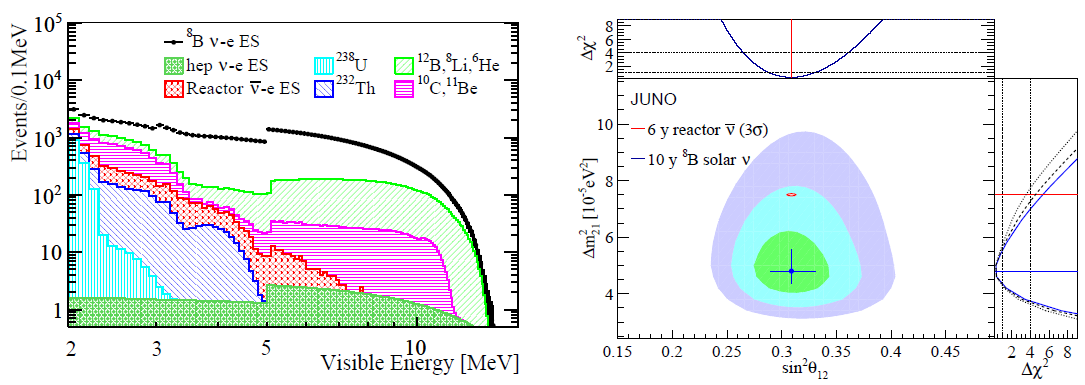}
\caption{Left: signal and backgrounds spectra above 2\,MeV, after solar analysis cuts and assuming 10 years exposure. Right: expected JUNO precision in the $\mathrm{sin^2} \theta_{12}$ and $\Delta m^2_{21}$ phase space, using $^8\mathrm{B}$ solar neutrinos and reactor antineutrinos. Figures from \cite{Solar_paper}.}
\label{fig:solar_ana}
\end{center}
\end{figure}

Another natural neutrino source accessible to JUNO is the Sun's core. As a result of the nuclear fusion processes occurring in our star, $\sim$MeV neutrinos can be detected by JUNO with unprecedented statistics, with respect to past experiments. Solar neutrinos are an important probe both to solar physics and to neutrino oscillation physics, two topics that will be addressed by JUNO \cite{Solar_paper}. Figure \ref{fig:solar_ana} shows the expected $^8$B solar neutrino signal and backgrounds in 10 years above 2\,MeV, which consists of an unprecedented limit for $^8$B neutrinos. The figure also reports the expected JUNO sensitivity with respect to the solar oscillation parameters, using both $^8$B neutrinos and reactor antineutrinos. This simultaneous measurement using two independent sources would be made for the first time.

The evaluation of the detector sensitivity towards lower--energy solar neutrinos, namely $pp$, $pep$, $^7$Be and CNO, is still in progress within the Collaboration, showing promising preliminary results.

\section{Atmospheric neutrinos}
\begin{figure}[htbp]
\begin{center}
\includegraphics[width=0.6\textwidth]{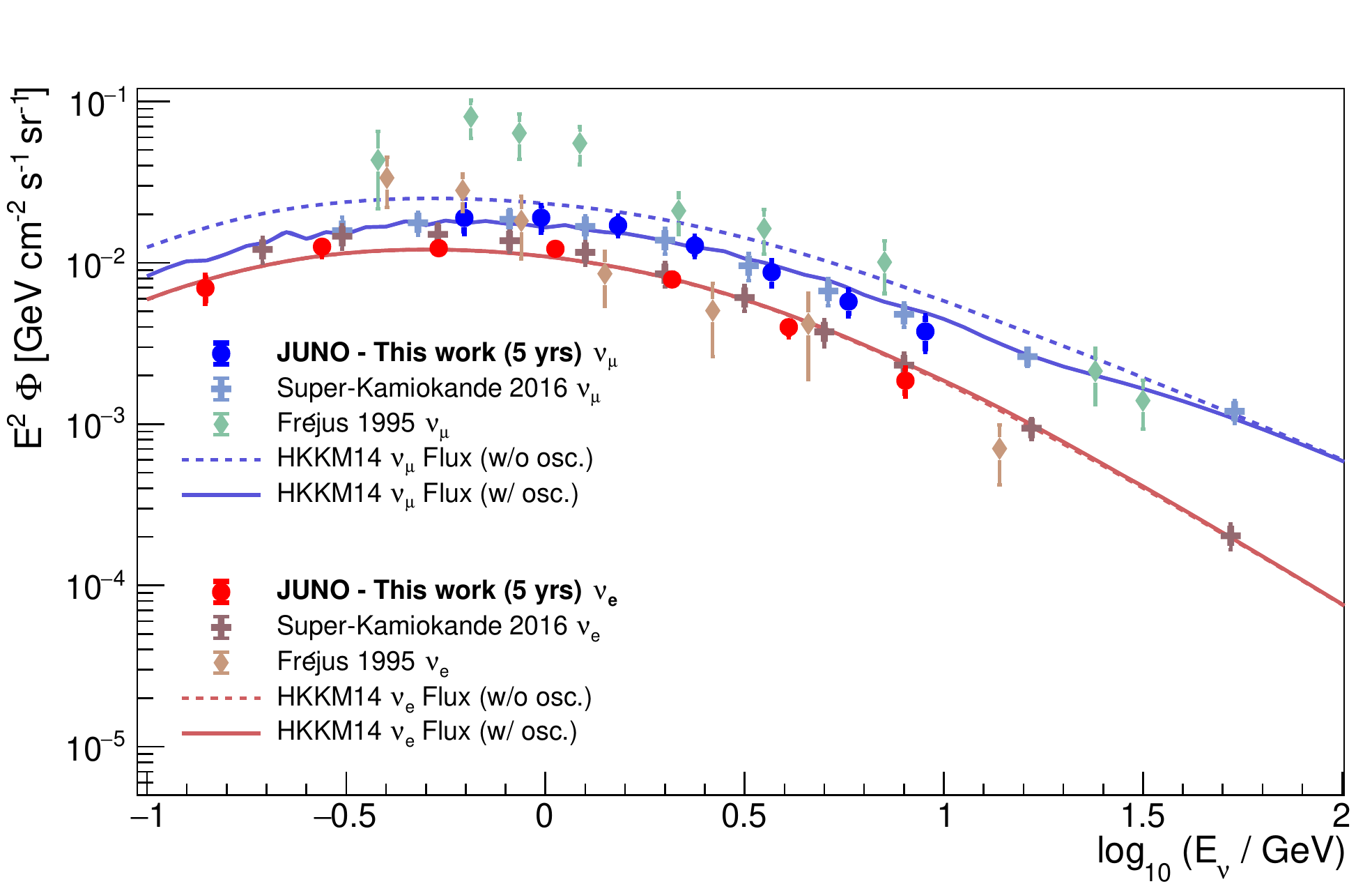}
\caption{Atmospheric neutrino energy spectra reconstructed by JUNO, assuming 5 years of detector exposure. Present measurements \cite{SK_atmo,Frejus_atmo} and HKKM14 \cite{HKKM14} model predictions in the same energy range are also reported. Figure from \cite{JUNO_NYB}.}
\label{fig:atmo_spectra}
\end{center}
\end{figure}

Neutrinos produced in the interactions of cosmic rays with the Earth's atmosphere can be also detected by JUNO, whose large size allows to reconstruct these events at the $\mathcal{O}$(GeV) energy range \cite{Atmo_discr,Atmo_paper}. The energy spectrum of atmospheric neutrinos can return many information about the production mechanisms. Moreover, it can help improving theoretical predictions in the multi--MeV region. Figure \ref{fig:atmo_spectra} shows the expected JUNO sensitivity towards the atmospheric $\nu_e$ and $\nu_\mu$ fluxes in the [0.1 - 10]\,GeV range, assuming an exposure of 5 years and the HKKM14 model \cite{HKKM14} as reference, also shown on the figure. Present measurements from other experiments are reported as well.

When crossing the Earth, atmospheric neutrinos can be also used to discriminate between the two MO scenarios, thanks to the matter effects acting during their propagation. Atmospheric neutrinos can thus be used by JUNO to provide a complementary measurement of neutrino MO, in a totally independent way with respect to reactor antineutrinos \cite{JUNO_YB}.

\section{Geoneutrinos}
\begin{figure}[htbp]
\begin{center}
\includegraphics[width=0.85\textwidth]{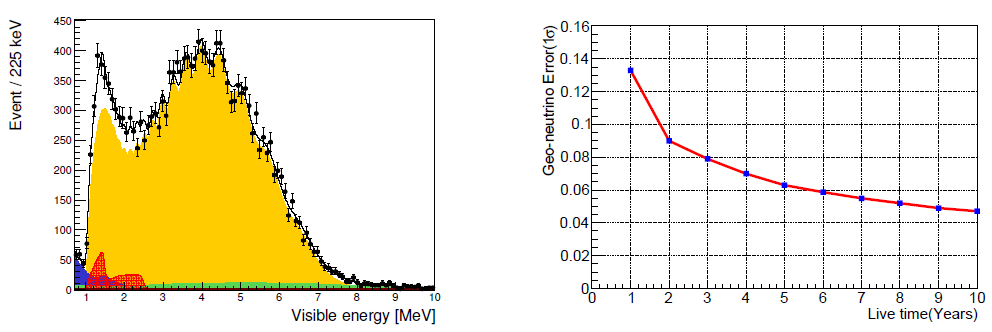}
\caption{Left: JUNO visible energy spectra of prompt Inverse Beta Decay candidates after geoneutrino analysis cuts, assuming 1 year of data--taking. Geoneutrino signal is in red, reactor antineutrinos  in orange, accidentals in blue, $^9$Li-$^8$He in green, $^{13}$C($\alpha$,$n$)$^{16}$O in magenta. Right: Expected 1$\sigma$ uncertainty on geoneutrino signal for JUNO as a function of live time, assuming a fixed chondritic Th/U ratio. Figures from \cite{JUNO_NYB}.}
\label{fig:geonu_spectrum}
\end{center}
\end{figure}

The JUNO size and energy resolution are also powerful features for the study of geoneutrinos at the $\mathcal{O}$(MeV) energy range. The ability of JUNO to reach a very high statistics of geoneutrino events within the first years of data taking opens exciting opportunities in probing the radiogenic content of the Earth and its distribution, improving current measurements. Today, indeed, there are still open issues regarding the net contribution of the Earth's mantle to the radiogenic power and the absolute Th/U ratio. JUNO will help to address these points. A  crucial input for the interpretation of results is the signal from the local crust, which has been studied by the Collaboration \cite{Crust_1,Crust_2}. As can be seen from Figure \ref{fig:geonu_spectrum}, the geoneutrino analysis has to deal with several backgrounds contribution, the most important being the reactor antineutrino flux from the nearby cores. In the same figure, the total uncertainty as a function of the live time is reported, to show that JUNO will be already competitive after 1 year of data--taking.

\section{Beyond the Standard Model searches}
\begin{figure}[htbp]
\begin{center}
\includegraphics[width=0.48\textwidth]{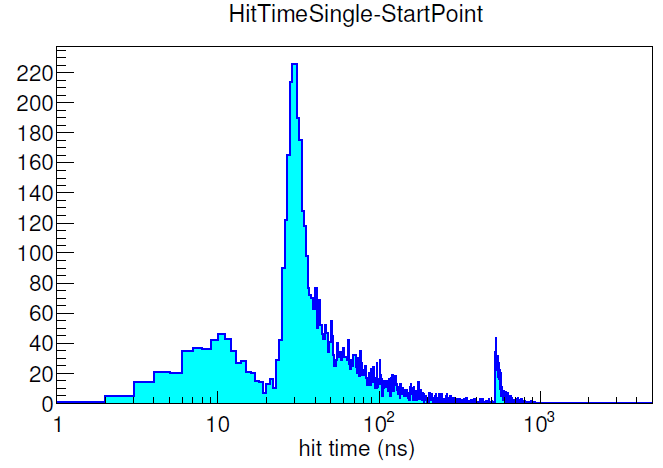}
\caption{Expected 3--fold coincidence signal at JUNO after a $p \rightarrow \overline{\nu} K^+$ event. The peaks from the $K^+$, the $\mu^+$ and the Michel electron are clearly visible. Figure from \cite{JUNO_NYB}.}
\label{fig:PD_time_spectrum}
\end{center}
\end{figure}

The search for exotic processes such as the proton decay will also be addressed by JUNO. In the $\mathcal{O}$(100\,MeV) range, a potential signal could be detected with high precision, with the possibility of setting the most stringent limit on the process. Figure \ref{fig:PD_time_spectrum} shows an example time distribution of a typical 3--fold coincidence of the $p \rightarrow \overline{\nu} K^+$ process, which JUNO would be able to detect with high efficiency.

Further potential topics which can be investigated by JUNO include the search of beyond-the-Standard-Model processes, regarding the nature of neutrinos themselves and Dark Matter, and the existence of light sterile neutrinos and magnetic monopoles.

\section{Conclusions}
The construction of the JUNO detector will be an exciting opportunity for the whole neutrino physics community, for many years to come. JUNO has been designed since the beginning as a state--of--the--art detector for MeV neutrino physics, but its impressive size and fine energy resolution pave the way to a wide range of measurements in the field of astroparticle and neutrino physics. Many interesting results are foreseen in the next years, cutting the edge in our knowledge of the Universe mechanisms at fundamental level.

\smallskip

\end{document}